\documentstyle[multicol,aps,prb,epsf]{revtex}

\newcommand{\Vec}[1]{{\bf #1}}

\begin{document}

\draft

\title{
Electronic structure of periodic curved surfaces\\
--- continuous surface versus graphitic sponge
}

\author{\noindent H. Aoki, M. Koshino, D, Takeda\cite{Takeda}, H. Morise}
\address{Department of Physics, University of Tokyo, Hongo, Tokyo
113-0033, Japan}
\author{K. Kuroki}
\address{Department of Applied Physics and
Chemistry, University of Electro-Communications, Chofu, 
Tokyo 182-8585, Japan}
\date{\today}

\maketitle

\begin{abstract}
We investigate the band structure of electrons
bound on periodic curved surfaces.  We have formulated 
Schr\"{o}dinger's equation with the 
Weierstrass representation when the surface is minimal, 
which is numerically solved.  Bands and the Bloch wavefunctions 
are basically determined by the way in which the ``pipes'' 
are connected into a network, where the Bonnet(conformal)-transformed 
surfaces have related electronic strucutres.  
We then examine, as a realisation of periodic surfaces, the 
tight-binding model for atomic networks (``sponges''), 
where the low-energy spectrum coincides with 
those for continuous curved surfaces.
\end{abstract}

\begin{multicols}{2}
\narrowtext

\section{Introduction}
While the physics of electron systems on flat two-dimensional planes 
(``2DEGs'') has been firmly established, it should be interesting to 
consider what will happen when we put electrons on 
{\it curved surfaces}.  
Here we consider electrons bound on periodic curved surfaces, 
envisaged as the ``crystal composed of surfaces''\cite{Koshino}.  
Geometrically, we can construct periodic curved surfaces 
that extend over the entire three-dimensional space
by connecting hyperbolic (i.e., everywhere negatively curved)
patches. Specifically, Schwarz has constructed  
{\it periodic minimal surfaces}, 
where minimal means that the negatively-curved 
surface has a minimized area\cite{Schwarz}.

In the materials-science viewpoint we can envisage that 
periodic minimal surfaces represent the topology 
of a class of condensed-matter systems, where 
atoms are arrayed along a curved surface.  
When the atomic sheet is graphitic, this corresponds to 
the carbon structure called C$_{60}$ zeolite or negative-curvature 
fullerene\cite{Mackay}, which are periodic surfaces 
if we smear out atoms in the effective-mass sense.   
It is then a fundamental question to consider how mobile (e.g., $\pi$)
electrons behave.   

In this paper, we first formulate
Schr\"{o}dinger's equation for free electrons on periodic surfaces, 
which turns out to be neatly expressed with Weierstrass's 
representation when the surface is minimal.  
We then numerically calculate the band structure for a typical 
periodic minimal surface, Schwarz's P-surface ($\sim$ C$_{60}$ 
zeolite; see Fig.\ref{fig:SurfaceUnit}). 

One interest is the relative importance of 
the effects of the global topology of the surface against the local atomic
configuration.  So we move on to a 
comparison of the band structure for a periodic continuous surface 
with that for an atomic network with a similar topology.  
We have found that their low-energy electronic structures are 
similar, i.e., primarily determined by 
how the wavefunction interferes on a multiply-connected surface.

\section{Weierstrass representation of minimal surfaces}
We start with a differential-geometric prerequisite for representing minimal
surfaces.   We consider a two-dimensional surface $\Vec{r}(u,v)$
embedded in 3D as parametrised by two coordinates $u,v$. 
According to Weierstrass and Enneper, 
when $\Vec{r}(u, v) = (x(u,v),y(u,v),z(u,v))$ is a minimal surface, 
there exist $F, G$, functions of $w \equiv u+iv$, 
with which $\Vec{r}$ is expressed as
\begin{eqnarray}
  {\bf r}(u, v) &=& {\rm Re} \left(
    \int^w_{w_0} F (1 - G^2) {\rm \, d}w,  \right. \nonumber \\
  && \left.
    \int^w_{w_0} i F (1 + G^2) {\rm \, d}w,
    \int^w_{w_0} 2F G {\rm \, d}w
  \right),
  \label{eqn:Weiermap}
\end{eqnarray}
where 
$FG^2$ is assumed to be regular: 
if not, we can exclude these points by incising cut(s) to make
$S\ni (u,v)$ a Riemann surface.

To construct Schr\"{o}dinger's equation for an electron on a curved surface, 
we adopt a physcial approach where we consider electrons bound 
to a thin, curved slab of thickness $d$, where
the limit $d\rightarrow 0$ is taken\cite{Nagaoka}.
With two-dimensional coordinates $(q^1,q^2)$ and metric tensor $g_{ij}$, 
the equation reads
    \begin{eqnarray}
      &&\left[  - {\hbar^2 \over 2m}
        {1 \over \sqrt{g}}
        {\partial \over \partial q^i} \sqrt{g}\ g^{ij}
        {\partial \over \partial q^j}
        - {\hbar^2 \over 8m}\left( \kappa_1 - \kappa_2 \right)^2
      \right] \psi(q^1, q^2) \nonumber \\ &&\hspace{45mm} = E\ \psi(q^1, q^2),
      \label{eqn:SchResult2}
    \end{eqnarray}
where summations over repeated indices are assumed,
and $\kappa_1,\kappa_2$ are the local principal curvatures.
An effect of the curvature appears as a curvature potential, 
$-(\hbar^2/8m)(\kappa_1 - \kappa_2)^2$, which is in general 
nonzero even when the surface is minimal ($\kappa_1+\kappa_2=0$ 
everywhere). 

If we apply Eq. (\ref{eqn:SchResult2}) to (\ref{eqn:Weiermap})
Schr\"{o}dinger's equation for a minimal surface
is neatly expressed in terms of $F$ and $G$ as
\begin{equation}
 -\frac{4}{|F|^2 (|G|^2 + 1)^2 }
      \left[
              {\partial^2 \over \partial u^2}
	      + {\partial^2 \over \partial v^2}
	      + \frac{4|G^\prime|^2}{(|G|^2 + 1)^2}
      \right] \psi = \varepsilon \ \psi,
\end{equation}
where $\varepsilon \equiv E/(\hbar^2/2m)$ and $G' = dG/dw$.
As seen from Eq. (\ref{eqn:Weiermap}),
$F$ has the dimension of length and $G$ is dimensionless.
Hence the energies in minimal surfaces always scale as 
$E/(\hbar^2/2mL^2)$, where 
$L \sim F \sim$ linear dimension of the unit cell.

We can have a more transparent form when $G(w)=w$ (as 
often the case with periodic minimal surfaces, including P-surface).  
We can then 
exploit the stereographic map (Gauss map) from the infinite 
complex plane $(u,v)$ to a unit sphere $(\theta, \phi)$
with $w=u+iv=\cot(\theta/2) {\rm e}^{i\phi}$, 
and we end up with a differential equation for $(\theta, \phi)$,
    \begin{eqnarray}
      -{ (1 - \cos\theta)^4 \over |F|^2 }
      \bigg(
              {\partial^2 \over \partial\theta^2}
              + \cot\theta{\partial \over \partial\theta}
               + {1 \over \sin^2 {\theta}}{\partial^2 \over
\partial\phi^2} \nonumber\\
             +  1\bigg) \psi  = \varepsilon \ \psi .
      \label{eqn:SchOnP}
    \end{eqnarray}
Curiously, a common coefficient $(1 - \cos\theta)^4/|F|^2$ factors out
for the Laplacian (the first line) 
and the curvature potential (+1). Hence the kinetic and potential
energies vary in a correlated manner as we go from one minimal surface
to another by changing $F$.  

\section{Results for Schwarz's P-surface}
\begin{figure}
\begin{center}
 \leavevmode\epsfxsize=80mm \epsfbox{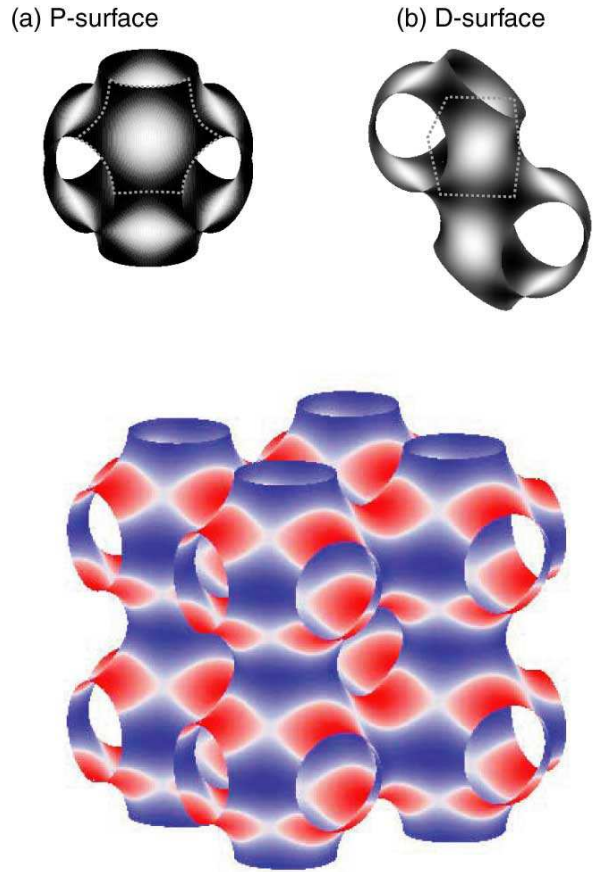}
\end{center}
\caption{
Above: the structures of P(a) and D(b) surfaces.  
The grey-scale represents the curvature potential.
Bottom: a typical wavefunction for P-surface.
}
\label{fig:SurfaceUnit}
\end{figure}

We first take Schwarz's P-surface
as a typical triply-periodic
(periodic in $x,y,z$) minimal surface, or a ``simple cubic network 
of pipes''. 
The Weierstrass-Enneper representation for the P-surface 
is\cite{Terrones} 
$F(w) = iL/\sqrt{1 - 14w^4 + w^8}$, 
where the unit cell size is $2.157L$.  
A unit cell of the P-surface comprises eight identical 
patches, as shown in Fig.\ref{fig:SurfaceUnit}.
With the stereographic mapping, a unit cell is 
mapped onto two spheres, connected
into a Riemann surface via four cuts, which have to be introduced to
take care of the poles in $F$.  

In Schr\"{o}dinger's equation for
periodic minimal surfaces, eqn.(\ref{eqn:SchOnP}), the variables $\theta,
\phi$ cannot be separated, so that we have solved the equation
numerically by discretising $\theta, \phi$ to diagonalise the
Hamiltonian matrix.  
The band structure is obtained by connecting the adjacent unit cells with
appropriate phase factors.

\begin{figure}
\begin{center}
 \leavevmode\epsfxsize=80mm \epsfbox{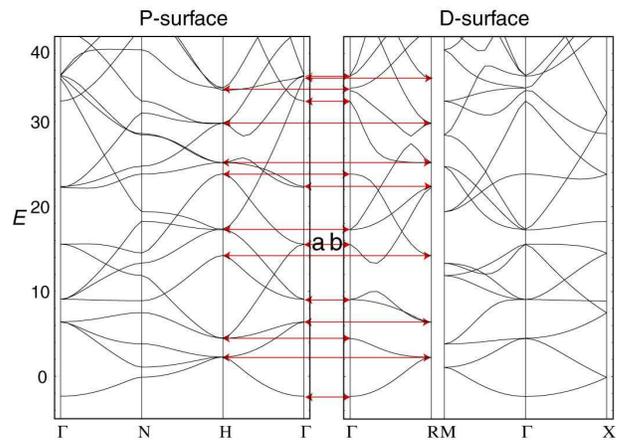}
\end{center}
\caption{
  The band structures for the P- and D-surfaces.
  Horizontal lines indicate how the energies at the zone center or
  edges coincide between the two cases.
  }
\label{fig:BandP_D}
\end{figure}

We now come to the result for the band structure for the P-surface in
Fig.\ref{fig:BandP_D} (left). 
The bands are displayed on the Brillouin zone for the bcc symmetry, 
since the P-surface happens to divide the space into two
equivalent parts where a body center enclosed by the 
surrounding unit cells has the same shape.  
In accord with the above argument, the energy scale (band width, splitting
etc) is $\sim \hbar^2/2mL^2$.  This is of the order of $1$ eV for 
$L\sim 10$ \AA, the unit cell size 
assumed for a hypothetical negative-curvature fullerene\cite{Townsend}.


\section{Bonnet transformation} 

One virtue of surfaces is we can deform one surface to another.  
For periodic minimal surfaces specifically, 
we can deform a minimal surface into another with 
the Bonnet transformation, a conformal transformation.  
Are the band structures for surfaces
connected by the Bonnet transformation related?  
A beautiful asset of the Weierstrass representation 
is that the Bonnet transformation is simply represented by 
a phase factor, $F \rightarrow Fe^{i\beta}$, in 
eqn.(\ref{eqn:Weiermap}), where $\beta$ is called the Bonnet angle.  
The P-surface (cubic) is Bonnet-transformed\cite{Terrones} 
into the G-surface (gyroid with $\beta=0.211\pi$) and
the D-surface (diamond with $\beta=\pi/2$; 
bottom panels in Fig.\ref{fig:SurfaceUnit}).

We can first note that, since the Bonnet transformation preserves 
the metric tensor and the Gaussian curvature, the
surfaces connected by a Bonnet transformation obey the identical 
Schr\"{o}dinger equation within a unit patch. This is readily seen in 
eqn.(\ref{eqn:SchOnP}), where $F$ only
enters as $|F|$, so $F\rightarrow Fe^{i\beta}$ does not alter the
equation.  However, this does not mean that the band
structures are identical, since the way in which unit cells are
connected is different among them.

Fig. \ref{fig:BandP_D} compares the band structures for the 
P- and D-surfaces.  While 
the band structures are indeed different, 
the band energy at {\it special
points} (Brillouin zone corners, edges and face-centers) have 
curiously identical
set of values between different surfaces.  
To be precise, the `law of correspondence' 
between P-surface $\Leftrightarrow$ D-surface is 
$\Gamma$, H $\Leftrightarrow$ $\Gamma$, R and 
N $\Leftrightarrow$ X, M, which can be shown from 
the phase factors connecting the different patches.

\section{Graphite sponges}
The band structure of atomic networks such as the C$_{60}$-zeolite 
or the negative curvature fullerene 
is expected to basically reflect the properties arising
from the curved surface on which the atoms are embedded, 
as far as the effective-mass formalism is applicable 
and effects of the ``frustration'' in the atomic configuration 
(odd-membered rings) are neglected.  
To see this, let us investigate the band structure of the 
graphite networks having a global topology similar 
to the periodic minimal surface.

\begin{figure}
\begin{center}
 \leavevmode\epsfxsize=80mm \epsfbox{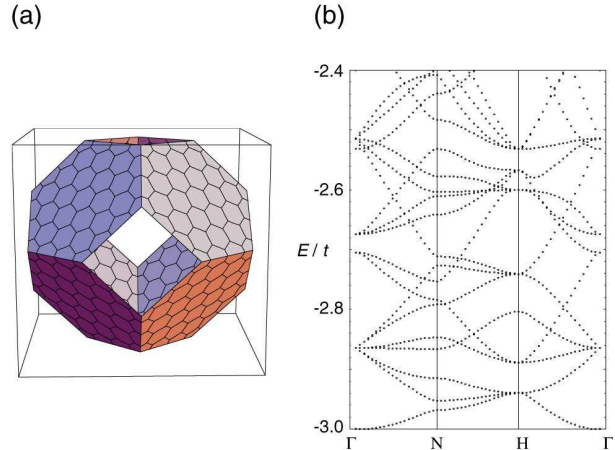}
\end{center}
\caption{
(a) Unit cell of P-surface-like graphite sponge, 
where each (octagonal) patch is a fragment of graphite.  
(b) Low-energy band structure for the graphite sponge, 
to be compared with Fig. \ref{fig:BandP_D} (left).
Unit of the energy is the hopping parameter $t$ 
and the origin the tight-binding band center.
}
\label{band_honey}
\end{figure}

Such ``sponges'', as termed by 
Fujita\cite{Fujita}, can be systematically constructed by 
connecting patches of atomic arrays (graphite fragments 
in the case of graphitic sponges).  
Among the families of the graphite sponges theoretically 
proposed by Fujita and coworkers\cite{Fujita}, 
here we take a sponge family as displayed in Fig. \ref{band_honey}(a) 
that is the closest realisation of the P-surface.
We have calculate its band structure 
in the single-band tight-binding model.

The results in Fig. \ref{band_honey}(b) shows that
the bands in a low-energy regime ($E < -2.5t$; $t$ the 
tight-binding hopping) has a one-to-one correspondence with 
those for the P-surface (Fig. \ref{fig:BandP_D}, left), 
which is natural since the effective-mass approach should be 
valid there.  
To be precise, there are
extra degeneracies in the sponge 
due to a symmetry in the atomic configuration.  
Naturally the correspondence with the continuous case 
is degraded for higher-energy regions, 
since the similarity holds only 
when the de-Broglie wavelength is greater than the atomic separation. 
Conversely, we can say that the electronic structure 
in the low-energy region is basically 
determined by the structure of the curved surface 
on which the atoms reside.

\section{Discussions}
We have formulated Schr\"{o}dinger's equation
for free electrons bound on periodic minimal surfaces, 
and obtained the band structures for P- and D-surfaces, 
which are Bonnet transformation of each other. 
While we have compared the 
tight-binding band for an atomic sponge network 
with those for the continuous periodic surface 
in the low-energy region, the relevant region is 
rather a higher ($E\simeq 0$) one.  For this, we have 
a zero-mass Dirac (i.e., Weyl) particle as far as the effective-mass 
picture for graphite is concerned.  So we should deal with 
a Dirac electron on periodic minimal surfaces, which 
should contain a richer physics, and the study is under way\cite{Fukuda}.

While we have taken only P- and D-surfaces,
G-surface, whose geometry is considered to be realised 
in a class of zeolite\cite{Gzeolite}, is distinguished from the other ones 
in that it has a triply chiral (i.e., spiral) symmetry.
It is known that a spiral symmetry makes bands to stick together, 
so it would be interesting to examine 
how the band structure in G-surface reflects this.

\end{multicols}
\end{document}